\documentclass[3p,twocolumn,showpacs,amsmath,amssymb,floatfix]{elsarticle2}
\usepackage[colorlinks=true]{hyperref}

\usepackage{mathpazo,times,graphicx}
\usepackage{mathtools}
\allowdisplaybreaks

\def\XXint#1#2#3{{\setbox0=\hbox{$#1{#2#3}{\int}$}
	\vcenter{\hbox{$#2#3$}}\kern-.5\wd0}}

\newcommand{\sech}{\mathrm{sech}}

\newcommand{\be}{\begin{equation}}
\newcommand{\ee}{\end{equation}}
\newcommand{\bse}{\begin{subequations}}
\newcommand{\ese}{\end{subequations}}
\newcommand{\bml}{\begin{multline}}
\newcommand{\eml}{\end{multline}}

\advance\textwidth 20mm
\advance\hoffset -10mm
\advance\textheight 28mm
\advance\voffset -14mm
\advance\parskip 2pt

\def\Real{\mathbb{R}}
\def\Complex{\mathbb{C}}
\numberwithin{equation}{section}
\def\Re{\mathop{\rm Re}\nolimits}
\def\Im{\mathop{\rm Im}\nolimits}

\journal{Physics Letters A}
\begin{document}
\begin{frontmatter}
\title{{\bf Linearizable boundary value problems for the nonlinear Schr\"odinger equation in laboratory coordinates}}
\author[rpi]{Katelyn Plaisier Leisman}

\author[buffalo]{Gino Biondini\corref{corespondingauthor}}
\cortext[correspondingauthor]{Corresponding author}
\ead{biondini@buffalo.edu}

\author[rpi]{Gregor Kovacic}

\address[rpi]{Rensselaer Polytechnic Institute, Department of Applied Mathematics, Troy, NY 12180, USA}
\address[buffalo]{State University of New York at Buffalo, Department of Mathematics, Buffalo, NY 14260, USA}
\begin{abstract}
Boundary value problems for the nonlinear Schr\"odinger equation on the half line in laboratory coordinates are
considered.  
A class of boundary conditions that lead to linearizable problems is 
identified by introducing appropriate extensions to initial-value problems
on the infinite line,
either explicitly or by constructing a suitable B\"acklund transformation.
Various soliton solutions are explicitly constructed and studied.
\end{abstract}

\begin{keyword}
Nonlinear Schr\"odinger equation, boundary value problems, laboratory frame, solitons
\\~
\end{keyword}
\end{frontmatter}


\section{Introduction}

Boundary value problems (BVPs) for integrable nonlinear wave equations have received renewed interest 
in recent years, 
thanks in part to the development of the so-called unified transform method, or Fokas method (e.g., see \cite{Fokas2008,FokasPelloni2014} and references therein).
One of the prototypical systems considered is the 
nonlinear Schr\"odinger equation (NLS), 
which we write as:
\be 
iq_T=q_{XX}+2|q|^2q.
\label{e:nlslc}
\ee
In many physical situations in which Eq.~\eqref{e:nlslc} arises as a description of the dynamics,
the solution $q(X,T)$ represents the slowly varying envelope describing the modulations of an underlying carrier wave.
In particular, when the NLS equation is obtained in the context of nonlinear optics, 
the independent variables $X$ and $T$ play the role of so-called light-cone variables,
which are related to coordinates in a laboratory frame by the transformation
$T=z_\mathrm{lab}$ and $X=t_\mathrm{lab}-z_\mathrm{lab}/c_g$
(modulo a suitable nondimensionalization),
where $z_\mathrm{lab}$ and $t_\mathrm{lab}$ are the physical spatial and temporal coordinates in the laboratory frame and 
$c_g$ is the speed of light in the specific nonlinear medium considered \cite{Agrawal,hasegawakodama,newellmoloney}.
A similar transformation to a comoving reference frame applies when Eq.~\eqref{e:nlslc} is derived as the governing equation for
the envelope of a wave train in deep water \cite{AS1981}.
(In that case, however, $X = x_\mathrm{lab} - c_g t_\mathrm{lab}$ and $T = t_\mathrm{lab}$, again modulo a suitable nondimensionalization.)

Equation~\eqref{e:nlslc} is of course a completely integrable system, and it possesses a rich mathematical structure.
In particular, it admits exact $N$-soliton solutions,
the simplest of which is the one-soliton solution:
\be
\kern-0.6em
q(X,T)=
A e^{i\left(V^2-A^2\right)T-iV X-i\psi}\mathrm{sech}\left[A \left(X-2V T-\delta\right)\right].
\label{e:nlslc1soliton}
\ee
Another aspect of integrability is the fact that the initial value problem (IVP) for Eq.~\eqref{e:nlslc} 
on the infinite line $-\infty<X<\infty$
can be solved using the inverse scattering transform (IST)
\cite{ZS1972} 
(see also~\cite{AS1981,NMZP1984,APT2004}).

Following the solution of the IVP for Eq.~\eqref{e:nlslc} in \cite{ZS1972}, 
the extension of the IST to BVPs was studied.
The special case of Eq.~\eqref{e:nlslc} on the half line $0<X<\infty$ with either homogeneous Dirichlet or Neumann boundary conditions (BCs)
[namely, $q(0,T)=0$ or $q_X(0,T)=0$, respectively] 
was studied in \cite{JMP16p1054} using the odd and even extension of the solution to the infinite line, respectively.
A general methodology applicable to a much broader class of BVPs was then recently presented in 
\cite{Fokas2008,FokasPelloni2014}.
One of the key results of the analysis is that, unlike the IVP, these BVPs can only be fully linearized
for a small subset of BCs.  
The BCs that give rise to linearizable BVPs are then called integrable, or alternatively \textit{linearizable} in the literature.
Note that calling such BCs linearizable may be confusing, since it is the BVP with the given BCs that is linearizable, not the BCs themselves.
Nonetheless, this terminology is widely used in the literature, and we will also use it in this work.

For the NLS Eq.~\eqref{e:nlslc}, the linearizable BCs are homogeneous Robin BCs, namely
\be
q_X(0,T) +\alpha q(0,T) =0,
\label{e:linearizable}
\ee
with $\alpha$ an arbitrary real constant.
Importantly, it was also shown in \cite{JPA24p2507,IP7p435,BH2009,SAM129p249} that, 
when such linearizable BCs are given, the BVP acquires extra symmetries.
In particular, using a nonlinear version of the method of images, it was shown in \cite{BH2009,SAM129p249} that
the discrete eigenvalues of the associated scattering problem appear in symmetric quartets, and that 
to each soliton in the physical domain $0<X<\infty$ there corresponds a mirror soliton in the virtual (reflected) domain $-\infty<X<0$.
This nonlinear method of images was then later extended and applied to other discrete and continuous NLS-type systems in \cite{BH2010,BB2015,caudrelier}.

However, since the coordinates $X$ and $T$ are relative to a reference frame moving with the group velocity of the carrier wave, 
a fixed boundary $X=0$ in the $XT$-frame corresponds to the line $z_\mathrm{lab}=c_gt_\mathrm{lab}$ in the laboratory frame.
Therefore, BVPs for Eq.~\eqref{e:nlslc} on the half line $0<X<\infty$ would correspond to 
BVPs with a boundary moving with the group velocity of the pulse in the laboratory frame,
which may not be the most useful ones to consider from a physical point of view.
(The above statement does not apply to Bose-Einstein condensates, where the NLS equation~\eqref{e:nlslc} arises without a transformation to a comoving reference frame.)

A more physically relevant set of problems to study is then perhaps those obtained by considering BVPs for the NLS equation in a laboratory frame,
namely, 
\vspace*{-1ex}
\be 
i\left(q_t+cq_x\right)=q_{xx}+2|q|^2q
\label{e:nls}
\ee
where $c$ is some relevant group velocity, 
and $x$ and $t$ are modified versions of $X$ and $T$, respectively,
without the transformation to a comoving frame.
For example, the additional group velocity term $cq_x$ arises when the NLS equation is derived in the context of both nonlinear optics 
\cite{Agrawal,hasegawakodama,newellmoloney}
and deep water waves
\cite{AS1981}.
For brevity, hereafter we refer to Eq.~\eqref{e:nls} as the ``NLSLab'' equation. 
The purpose of this Letter is precisely to study BVPs for Eq.~\eqref{e:nls} on the half line $0<x<\infty$.  
In particular, we identify a class of linearizable BCs which reduces to Eq.~\eqref{e:linearizable} for $c=0$,
we construct a class of exact soliton solutions of the BVP and we study the corresponding behavior.

Of course Eq.~\eqref{e:nls} is also completely integrable.
[Indeed, the additional term simply corresponds to the addition of one of the lower flows in the integrable hierarchy associated with Eq.~\eqref{e:nlslc}.]
Therefore, one can expect to still be able to use a suitably modified implementation of the IST to study both the IVP and the BVP.
On the other hand, we will see that some of the properties of Eq.~\eqref{e:nls} differ significantly from those of Eq.~\eqref{e:nlslc}.
In particular, the linearizable BCs are different, and the behavior of solutions is also somewhat different.

The outline of this work is the following.
In section~\ref{s:IVP}
we give a brief overview of the IST for the NLSLab Eq.~\eqref{e:nls} on the infinite line, which will be used later on to study BVPs. 
The direct and inverse scattering problems are the same as for the NLS equation~\eqref{e:nlslc} in the light-cone frame, 
but the evolution problem has an additional term.  
In section~\ref{s:extension} we begin to study BVPs for Eq.~\eqref{e:nls},
and in particular we show how a modified extension to a suitable problem the infinite line allows one to 
identify a class of linearizable BCs for the NLSLab equation on the half line.  
Then we use these results 
to solve the BVP for the NLSLab equation with those linearizable BCs.
In particular, we show that, 
like with Eq.~\eqref{e:nlslc}, discrete eigenvalues in the BVP appear in symmetric quartets.
In this case, however, instead of being symmetric about the origin in the spectral plane, 
they are symmetric about the point $k=-c/4$.  
In section~\ref{s:BT} we construct a different extension based on a certain B\"acklund transformation,
and we show that additional linearizable BCs can be obtained in this way.
In section~\ref{s:solitons}
we construct explicit soliton solutions to the BVP and discuss their behavior.
In particular, we show that, as a result, solitons are still reflected from the boundary $x=0$.  
Finally, in section~\ref{s:conclusions} we offer some concluding remarks.


\section{The NLSLab equation on the infinite line} 
\label{s:IVP}

The NLSLab Eq.~\eqref{e:nls} is the compatibility condition of the matrix Lax pair
\begin{subequations}
\begin{align}
\phi_x&= (-ik\sigma_3+Q)\phi\label{e:laxs}\,,
\\
\phi_t&=H\phi-c(-ik\sigma_3+Q)\phi\,.
\label{e:laxe}
\end{align}\label{e:lax}
\end{subequations}
where 
\bse 
\begin{align} Q(x,t) &= \left[\begin{array}{cc}0&q\\r&0\end{array}\right], \hskip 20pt \sigma_3=\left[\begin{array}{cc}1&0\\0&-1\end{array}\right]
\\
H(x,t,k) &= 2ik^2\sigma_3 +iQQ\sigma_3+ iQ_x\sigma_3-2kQ\end{align}
\ese
and with the symmetry $r= - q^*$ as usual for the focusing case. 
The scattering problem \eqref{e:laxs} coincides with that for the NLS equation in the light-cone frame.
Thus, the direct and inverse scattering problems are exactly the same as usual.  
Correspondingly, we limit ourselves to only reviewing the essential points of the direct and inverse problem,
referring the reader to \cite{AS1981,APT2004} for all details.
If one takes $c=0$, the formalism reduces to the IST for the NLS equation in the light-cone frame.

\paragraph*{Direct scattering problem}
The direct problem consists of constructing $k$-dependent scattering data from the potential $q(x,t)$
of the scattering problem.  

For all $k\in\Real$, 
the scattering problem~\eqref{e:laxs} admits Jost eigenfunctions with asymptotic behavior
\be
\phi^{\pm}(x,t,k) = e^{-ikx\sigma_3} + o(1),\qquad x\to\pm\infty.
\label{e:bcs}
\ee
One then introduces modified Jost solutions with constant BCs at infinity as 
\be
\mu^{\pm}(x,t,k) = \phi^{\pm}(x,t,k)e^{+ikx\sigma_3},
\label{e:modjost}
\ee
such that
\bse
\label{e:jostode}
\begin{gather}
\mu_x^{\pm}+ik[\sigma_3,\mu^{\pm}]=Q\mu^{\pm}\,,
\\
\mu^\pm \to I\,,\qquad x\to \pm\infty\,.
\end{gather}
\ese
The problem~\eqref{e:jostode} is converted into Volterra integral equations using the integrating factors 
$\psi^{\pm}=e^{+ikx\sigma_3 }\mu^{\pm}e^{-ikx\sigma_3}$, to obtain
\bse
\label{e:jostint}
\begin{align}
\mu^+&(x,t,k) = I 
\nonumber\\
  &{ } -\int_{x}^{\infty} e^{-ik\sigma_3(x-s)}Q(s,t) \mu^{\pm}(s,t,k)e^{+ik\sigma_3(x-s)}ds\,,
\\
\mu^-&(x,t,k) = I 
\nonumber\\
  &{ } + \int_{-\infty}^{x} e^{-ik\sigma_3(x-s)}Q(s,t) \mu^{\pm}(s,t,k)e^{+ik\sigma_3(x-s)}ds\,.
\end{align}
\ese
Introducing the notation $\mu^\pm = (\mu_1,\mu_2)$ to denote the columns of $\mu^\pm$, one then sees that 
$\mu_1^-$ and $\mu_2^+$ can be analytically extended in the upper-half plane (UHP) [i.e., $\Complex^+ = \{k\in\Complex:\Im k>0\}$], 
and $\mu_1^+$ and $\mu_2^-$ in the LHP.  

%

Since both $\phi^\pm$ are fundamental matrix solutions of Eq.~\eqref{e:laxs} $\forall k\in\Real$,
one can introduce a scattering matrix as 
\begin{gather}
\phi^+(x,t,k) = \phi^-(x,t,k)S(k,t),\quad k\in\Real,
\label{e:scat}
\\
S(k,t) = \left[\begin{array}{cc}s_{11}&s_{12}\\ s_{21}&s_{22}\end{array}\right].
\end{gather}
%
%
One can then show that $s_{11}$ can also be analytically extended in LHP, and $s_{22}$ in UHP.  
We can rewrite Eq.~\eqref{e:scat} as a columnwise relation in terms of the reflection coefficients 
$b(k,t) = {s_{21}}/{s_{11}}$ and $\tilde b(k,t) = {s_{12}}/{s_{22}}$ as
\bse
\label{e:scatrefl}
\begin{gather}
\frac{\mu_1^+(x,t,k)}{s_{11}(k,t)} = \mu_1^-(x,t,k) + e^{+2ikx} b(k,t)\mu_2^-(x,t,k),
\\
\frac{\mu_2^+(x,t,k)}{s_{22}(k,t)} = e^{-2ikx}\tilde b(k,t) \mu_1^-(x,t,k) + \mu_2^-(x,t,k),
\end{gather}
\ese

The proper eigenvalues of the scattering problem are the values $k\in\Complex\setminus\Real$ 
such that the corresponding solution to the scattering problem vanishes as $|x|\to\infty$.  
These occur for $k_j\in\Complex^-$ when $s_{11}(k_j,t)=0$, so that $\phi_1^+(x,t,k_j)=c_j(t)\phi_2^-(x,t,k_j)$, 
with norming constant $c_j(t)$, 
and for $\tilde k_j\in\Complex^+$ when $s_{22}(\tilde k_j,t)=0$, so that $\phi_2^+(x,t,\tilde k_j)=\tilde c_j(t)\phi_1^-(x,t,\tilde k_j)$, 
with norming constant $\tilde c_j(t)$.

The symmetry $r=-q^*$ of the scattering problem (which yields the skew-Hermitian condition $Q=-Q^\dagger$) implies that 
if $\phi(x,t,k)$ is a solution, $[\phi^\dagger\left(x,t,k^*\right)]^{-1}$ is also.  
As a result, one has
$\left(\phi^\pm(x,t,k)\right)^{-1} = \left(\phi^\pm(x,t,k)\right)^\dagger$ $\forall k\in\Real$,
which yields
$S^{-1}(k,t) = S^\dagger(k,t)$ $\forall k\in\mathbb{R}$.
In turn, this yields
\bse\begin{align}
&s_{11}^*(k^*,t) = s_{22}(k,t),\quad \Im k>0,\\
&s_{21}^*(k,t) = -s_{12}(k,t),\quad k\in\Real,\\
&\tilde b(k,t) = -b^*(k,t),\quad k\in\Real,\\
&\tilde c_j(t) = -c_j^*(t),\\
&\tilde k_j=k_j^*.
\label{e:pairs}
\end{align}
\ese
In particular, Eq.~\eqref{e:pairs} implies that the proper eigenvalues appear in complex conjugate pairs.

\paragraph*{Inverse scattering problem}
The inverse problem consists of constructing a map from the scattering data
(namely, the reflection coefficients, proper eigenvalues, and norming constants) back to the potential $q(x,t)$. 
%
%
This is done by 
constructing a suitable matrix Riemann-Hilbert problem (RHP).  
Namely, taking into account 
the analyticity properties of the eigenfunctions and scattering coefficients, 
one defines sectionally meromorphic functions
\bse 
\label{e:mero}
\begin{gather}
M_-(x,t,k) = \left[\frac{\mu_1^+}{s_{11}},\mu_2^-\right],\quad \Im k\ge0\,,
\\
M_+(x,t,k) = \left[\mu_1^- , \frac{\mu_2^+}{s_{22}}\right],\quad \Im k\le0\,.
\end{gather}
\ese
Then, in light of Eq.~\eqref{e:mero}, Eq.~\eqref{e:scatrefl} yields the jump condition for the RHP as 
\be 
M_-(x,t,k) = M_+(x,t,k) \big[ I + R(x,t,k) \big],\quad k\in\Real\,,
\label{e:rhp}
\ee
where the jump matrix is
\be R(x,t,k) = \left[\begin{array}{cc}-b(k,t)\tilde b(k,t) & -\tilde b(k,t)e^{-2ikx}\\ b(k,t)e^{+2ikx} & 0\end{array}\right]\label{e:jump}\ee

Since $M_{\pm}(x,t,k)=I+O(1/k)$ as $k\to\infty$ and is analytic except for poles at $k_j$ and $\tilde k_j$, 
after subtracting the residue contributions from these poles one can apply Cauchy projectors 
to Eq.~\eqref{e:rhp},
obtaining formally the solution of the RHP as 
\bml 
M_\pm(x,t,k) = I +\sum_{j=1}^J\frac{\mathrm{Res}[M_-,k_j]}{k-k_j}+\sum_{j=1}^{\tilde J}\frac{\mathrm{Res}[M_+,\tilde k_j]}{k-\tilde k_j}\\-\frac{1}{2\pi i}\int \frac{M_+(x,t,\zeta)R(x,t,\zeta)}{\zeta-(k\pm i0)}d\zeta
\label{e:meroeq}.
\end{multline}


From the asymptotic behavior of Eq.~\eqref{e:jostint} as $k\to\infty$, 
one obtains $q(x,t)=\lim_{k\to\infty} -2i\mu_{12}(x,t,k)$.  
Evaluating the asymptotic behavior of Eq.~\eqref{e:meroeq} as $k\to\infty$ and comparing, 
one can then finally reconstruct the solution of the NLS equation~\eqref{e:nls} as 
%
%
\begin{multline}
q (x,t) = -2i\sum_{j=1}^J\tilde C_j(t)e^{-2i\tilde k_jx}\mu_{11}^- (x,t,\tilde k_j)\\+\frac{1}{\pi}\int\tilde b(\zeta,t)e^{-2i\zeta x}\mu_{11}^- (x,t,\zeta)d\zeta,
\label{e:qnoev}
\end{multline}
with
\bse
\begin{align}
&\begin{multlined}[b][.40\textwidth] 
\mu_2^-(x,t,k) = \begin{pmatrix}0\\1\end{pmatrix} +\sum_{j=1}^{\tilde J}\frac{\tilde C_j(t) e^{-2i\tilde k_jx}\mu_1^- (x,t,\tilde k_j)}{k-\tilde k_j}\\+\frac{1}{2\pi i}\int \frac{\tilde b(\zeta,t)e^{-2i\zeta x}\mu_1^- (x,t,\zeta)}{\zeta-(k-i0)}d\zeta,
\end{multlined}\\
&\begin{multlined}[b][.40\textwidth] 
\mu_1^-(x,t,k) = \begin{pmatrix}1\\0\end{pmatrix} +\sum_{j=1}^J\frac{C_j(t)e^{+2ik_jx}\mu_2^- (x,t,k_j)}{k-k_j}\\-\frac{1}{2\pi i}\int \frac{ b(\zeta,t)e^{+2i\zeta x}\mu_2^- (x,t,\zeta)}{\zeta-(k+i0)}d\zeta,
\end{multlined}
\end{align}\label{e:munoev}\ese
where $C_j(t) = {c_j(t)}/{s_{11}'(k_j)}$ and $\tilde C_j(t)= {\tilde c_j(t)}/{s_{22}'(\tilde k_j)}$, 
and primes denote derivative with respect to $k$.  
Note that the symmetry $r = - q^*$ implies $\tilde C_j(t)=-C_j^*(t)$.

\paragraph*{Time evolution}
The last step in the implementation of the IST is to compute the time evolution of the norming constants and reflection coefficients using the second half [Eq.~\eqref{e:laxe}] of the Lax pair.  
Here is where the IST for the NLSLab equation~\eqref{e:nls} differs from that for the light-cone NLS equation~\eqref{e:nlslc}. 

Let $\Phi(x,t,k)$ be a simultaneous solution of both parts of the Lax pair.
Assuming $q$ and $q_x$ both vanish as $x\to\pm\infty$, 
Eq.~\eqref{e:laxe} yields
$\Phi_t = i(kc+2k^2)\sigma_3\,\Phi + o(1)$ as $x\to \pm \infty$,
implying
\be
\Phi(x,t,k) = e^{i(kc+2k^2)t\sigma_3}\,\Phi(x,0,k) + o(1)\,\quad x\to\pm\infty\,.
\label{e:phitime}
\ee
The Jost solutions $\phi_\pm(x,t,k)$ of the scattering problem satisfy the fixed BCs~\eqref{e:bcs},
which are not compatible with this time evolution.
On the other hand, since both $\phi_\pm(x,t,k)$ are fundamental matrix solutions of the scattering problem,
we can express $\Phi(x,t,k)$ in terms of them
as
$\Phi_\pm(x,t,k) = \phi_\pm(x,t,k)\,\chi_\pm(k,t)$ $\forall x,k\in\Real$.
Comparing with Eq.~\eqref{e:phitime} we then have 
\be
\Phi^\pm (x,t,k) = 
\phi^\pm (x,t,k)\,e^{i(kc+2k^2)t\sigma_3}\,\qquad x,k\in\Real\,.
\label{e:truetime}
\ee
Conversely, Eqs.~\eqref{e:laxe} and~\eqref{e:truetime} yield the time evolution of the Jost solutions as
%
%
%
\be
\phi_t^\pm = [H-c(-ik\sigma_3+Q)]\,\phi^\pm - i(kc+2k^2)\phi^\pm\sigma_3\,.
\label{e:evphip}
\ee
Finally, combining Eqs.~\eqref{e:evphip} and~\eqref{e:scat} 
%
%
we obtain an evolution equation for the scattering matrix as 
\be
S_t = (ikc+2ik^2)\left[\sigma_3,S\right],\quad k\in\Real
\ee
(where $[\cdot,\cdot]$ denotes matrix commutator),
which we can solve to find the dependence of the norming constants and reflection coefficients:
%
%
\bse
\label{e:evbc}
\begin{align}
b(k,t) &= e^{-2i(kc+2k^2)t}b(k,0),\quad k\in\Real,
\\
C_j(t) &= e^{-2i(k_jc+2k_j^2)t}C_j(0)\,.
\end{align}
\ese

\paragraph*{Soliton solutions}
%
%
In the simplest case of a single discrete eigenvalue without radiation one obtains the one-soliton solution
of the NLSLab equation.
That is, taking $b(k,0)=0~\forall k\in\Real$, $J=1$, $k_1=(V+iA)/2$ and $C_1(0)=Ae^{A\delta+i(\psi+\pi/2)}$
we have
\begin{multline}
q (x,t) = A\,e^{-iVx + i(V^2-A^2)t+iVct-i\psi}
\\ 
\times \sech\left[A\left(x-\left(2V+c\right)t - \delta\right)\right].
\label{e:soliton}
\end{multline}
Compared to the one-soliton solution of the NLS equation in light-cone variables, note the 
modified relation between the discrete eigenvalue and the soliton velocity and 
appearance of an additional time-dependent phase factor.

Like with the NLS equation in the light-cone frame, 
one can also derive a determinantal expression for more general $N$-soliton solutions,
and one can directly obtain the analytic scattering coefficient from the inverse problem via the trace formulae:
\vspace*{-1ex}
\be 
s_{11}(k)=\exp\bigg[-\frac1{2\pi i}\int_{-\infty}^\infty \frac{\log(1+|b(\zeta)|^2)}{\zeta-k}d\zeta\bigg]
  \prod_{j=1}^J\frac{k-k_j}{k-k_j^*}.
\label{e:trace}
\ee
Also, in what follows we found it convenient to parametrize the discrete eigenvalues as
$k_j=(V_j+iA_j)/2$
and the norming constants as
\be 
C_j(0)=iA_j\,e^{A_j\delta_j+i\psi_j}\prod_{\genfrac{}{}{0pt}{}{\ell=1}{\ell\ne j}}^J\frac{k_j-k_\ell^*}{k_j-k_\ell}.
\label{e:normingconst}
\ee
The product in the right-hand side of Eq.~\eqref{e:normingconst} is nothing but 
the value of $1/s_{11}'(k_j)$ in the reflectionless case, as obtained from the trace formulae, 
and will make some key formulae for the BVP much simpler than the corresponding ones in \cite{BH2009,SAM129p249}
(cf. section~\ref{s:solitons}).

\section{Linearizable BCs for the NLSLab equation}
\label{s:extension}

\subsection{Explicit extension to the infinite line}

As mentioned earlier, the BVP for the NLS equation on the half line in light-cone variables was studied in \cite{JMP16p1054,BH2009} 
in the case of homogeneous Dirichlet or Neumann BC
using respectively an odd or even extension of the potential to the infinite line.  
The use of either of these extensions was possible because the light-cone NLS is invariant to the transformation $x\mapsto -x$.  
The NLSLab equation~\eqref{e:nls} does not possess this invariance, however, so the odd and even extensions are not useful for the BVP, and 
one is forced to use a different extension.  
In particular, here we will use the following modified extension:
\bml
\kern-0.6em
\hat Q (x,t) = Q (x,t)\Theta(x)\\-e^{i(\gamma+cx/2)\sigma_3}Q(-x,t)e^{-i(\gamma+cx/2)\sigma_3}\Theta(-x),
\label{e:ext}
\end{multline}
where $\gamma\in\Real$ is an arbitrary constant and 
$\Theta(x)$ is the Heaviside theta function [namely, $\Theta(x)=1$ for $x\ge0$ and $\Theta(x)=0$ for $x<0$].  
When $c=0$, taking $\gamma=0$ yields the odd extension of the potential and $\gamma=\pi/2$ the even extension.
Owing to the periodicity of the exponential function, we can restrict ourselves to taking $\gamma\in[0,\pi)$
without loss of generality. 

In section~\ref{s:validity} we show that, for $\gamma=0,\pi/2$, the extended potential $\hat Q(x,t)$ 
is continuous and differentiable for all $x\in\Real$, including at $x=0$, 
and that it satisfies the NLSLab equation~\eqref{e:nls} for all $x\in\Real$,
including at $x=0$.
Moreover, for these values of $\gamma$ the extended potential satisfies nontrivial BCs.
Specifically, taking $\gamma=0$ yields
\be
\hat q(x,t) = q(x,t)\Theta(x) - e^{icx} q(-x,t)\Theta(-x),
\label{e:Dext}
\ee
and the potential satisfies homogeneous Dirichlet BC, namely,
\be 
q(0,t)=0\,.
\label{e:DBC}
\ee
Conversely, taking $\gamma=\pi/2$ yields
\be
\hat q(x,t) = q(x,t)\Theta(x) + e^{icx} q(-x,t)\Theta(-x),
\label{e:Rext}
\ee
and the potential satisfies the following Robin BC,
\be 
2iq_x(0,t) + c\,q(0,t)=0,
\label{e:RBC}
\ee
which reduces to a standard homogeneous Neumann BC when $c=0$.  
In section~\ref{s:bvpsoln} we will then show how, 
when either Eq.~\eqref{e:DBC} or~\eqref{e:RBC} are given, 
the corresponding BVP for the NLSLab equation~\eqref{e:nls} is completely linearized 
via the extension~\eqref{e:Dext} or~\eqref{e:Rext}, respectively.
Thus, Eqs.~\eqref{e:DBC} and~\eqref{e:RBC} identify
two classes of linearizable BCs for the NLSLab equation on the half line.

Importantly, $\gamma=0,\pi/2$ are the only values that result in nontrivial BCs at $x=0$.
This is because, when $\gamma\ne n\pi/2$, continuity of the extended potential at $x=0$ requires $q(0,t)=q_x(0,t)=0$, 
which yields the trivial solution $q (x,t)\equiv0$.
In fact, we also show below that, 
if one considers extensions of the form 
\be 
\hat Q (x,t) = Q (x,t)\Theta(x)+F (x,t)Q(-x,t)G (x,t)\Theta(-x),
\label{e:generalext}
\ee
\eqref{e:ext} is the only member in such class that satisfies Eq.~\eqref{e:nls}.  
Since the extension~\eqref{e:ext} never satisfies Neumann BCs, it follows that the BVP for the NLSLab equation 
with Neumann BCs cannot be be solved using this approach.

\let\==\bar

\subsection{Validity and uniqueness of the extension}
\label{s:validity}

We now demonstrate that the extension defined by Eq.~\eqref{e:ext}
satisfies the NLSLab equation~\eqref{e:nls} 
for all $x\in\Real$ and that in addition it satisfies appropriate BCs at $x=0$.  
We also demonstrate that, among the more general family of extensions~\eqref{e:generalext},
only Eq.~\eqref{e:ext} has these properties.

Consider the general extension~\eqref{e:generalext}.
Obviously Eq.~\eqref{e:ext} satisfies the NLSLab equation for $x>0$.
Accordingly, we just need to focus on the cases $x<0$ and $x=0$.  
To begin with, note that,
for $x<0$, we have
\be 
\hat Q(x,t)=-\left[\begin{array}{cc}f_{11}g_{21}\=q+f_{12}g_{11}\=r &f_{11}g_{22}\=q+f_{12}g_{12}\=r\\ 
  f_{21}g_{21}\=q+f_{22}g_{11}\=r &f_{21}g_{22}\=q+f_{22}g_{12}\=r
\end{array}\right],
\nonumber
\ee
where $f_{ij}$ and $g_{ij}$ denote the corresponding matrix entries of $F$ and $G$
and where for brevity we denoted $\=q=q(-x,t)$ and $\=r=r(-x,t)$.
Since $\hat Q$ must be off-diagonal, 
we need
\bse
\be 
f_{11}g_{21}=f_{12}g_{11}=f_{21}g_{22}=f_{22}g_{12}=0.
\label{e:condition1}
\ee
Moreover, requiring that $\hat Q$ be nontrivial and that the equations for $\hat q$ and $\hat r$ be compatible, 
we also get
\be 
f_{12}g_{12}=f_{21}g_{21}=0, \quad
f_{11}g_{22}\ne0, \quad 
f_{22}g_{11}\ne0.
\label{e:condition3}
\ee
\ese
Combining conditions~\eqref{e:condition1} and~\eqref{e:condition3}, we obtain $f_{12}=f_{21}=g_{12}=g_{21}=0$, 
implying that $F$ and $G$ must be diagonal matrices.  
So, the extension~\eqref{e:generalext} for $x<0$ reduces to
\be 
\hat Q(x,t)=-\left[\begin{array}{cc}0&f_{11}g_{22}q(-x,t)\\ f_{22}g_{11}r(-x,t)&0\end{array}\right],
\ee
or simply, in component form,
\bse
\label{e:reducedextension}
\begin{align}
\hat q(x,t)&=-f_{1}(x,t)g_{2}(x,t)q(-x,t)\\ 
\hat r(x,t)&=-f_{2}(x,t)g_{1}(x,t)r(-x,t)
\end{align}
\ese
for $x<0$.

Next we check the compatibility of the above extension with the NLSLab equation for $x<0$. 
%
Substituting Eq.~\eqref{e:reducedextension} into Eq.~\eqref{e:nls},
the equation for $\hat q$ becomes
\bml 
q\left[i(f_1g_2)_t+ic(f_1g_2)_x-(f_1g_2)_{xx}\right]
\\
=2|q|^2q\left[(f_1g_2)(|f_1g_2|^2-1)\right]
\\
+2q_x\left[ic(f_1g_2)-(f_1g_2)_x\right],
\end{multline}
where we simplified the result by making use of the fact that $q(x,t)$ satisfies the NLS equation.
To avoid imposing unnecessary restrictions on $q$, we therefore must have 
\bse
\be 
|f_1g_2|^2=1,\quad (f_1g_2)_x=ic(f_1g_2),\quad (f_1g_2)_t=0.
\ee
Similar considerations based on the equation for $\hat r$ (with $\hat r=-\hat q^*$ as usual) indicate that we also need
\be 
|f_2g_1|^2=1,\hskip 10pt (f_2g_1)_x=-ic(f_2g_1),\hskip 10pt (f_2g_1)_t=0.
\ee
\ese
There are a number of mathematically equivalent solutions to this problem.
On the other hand, without loss of generality, we choose
\be 
F(x)=e^{i\sigma_3(\gamma+cx/2)},\hskip 20pt G(x)=e^{-i\sigma_3(\gamma+cx/2)}.
\ee
with $\gamma\in[0,\pi)$.
Thus, we have proved that the extension~\eqref{e:ext} is unique in its ability to generate a solution of the NLSLab equation 
for both $x>0$ and $x<0$.  

It therefore only remains to show is that the extended field $\hat q(x,t)$ also solves the NLSLab equation at $x=0$.
To this end, it is sufficient to prove that $\hat Q(x,t)$, $\hat Q_t(x,t)$, $\hat Q_x(x,t)$, and $\hat Q_{xx}(x,t)$ are all continuous at $x=0$.
That is, it is sufficient to show that the following conditions hold:%
\bse
\label{e:continuity}
\begin{align}
\lim_{x\rightarrow0^+} \hat Q(x,t) &= \lim_{x\rightarrow0^-}\hat Q(x,t),
\label{e:continuity0}
\\
\lim_{x\rightarrow0^+} \hat Q_t(x,t) &= \lim_{x\rightarrow0^-}\hat Q_t(x,t),
\label{e:continuity1}
\\
\lim_{x\rightarrow0^+} \hat Q_x(x,t) &= \lim_{x\rightarrow0^-}\hat Q_x(x,t),
\label{e:continuity2}
\\
\lim_{x\rightarrow0^+} \hat Q_{xx}(x,t) &= \lim_{x\rightarrow0^-}\hat Q_{xx}(x,t).
\label{e:continuity3}
\end{align}
\ese
Note that Eqs.~\eqref{e:continuity} are sufficient but not necessary, 
and it could be possible for $\hat q(x,t)$ to satisfy the NLSLab equation at $x=0$ 
even if $\hat q$, $\hat q_x$, $\hat q_{xx}$ and $\hat q_t$ are all individually discontinuous there.
We will also show in the next section that other linearizable BC can be obtained with alternative methods.

It should be clear that, because of the definition we have chosen for the Heaviside function, 
each of the above quantities is continuous from the left.
It should also be clear that, when the continuity conditions~\eqref{e:continuity} are all satisfied,
the extension $\hat Q(x,t)$ also satisfies the NLSLab equation at $x=0$, 
since $Q(x,t)$ satisfies the same equation from the right.
Next we study each of the conditions in Eq.~\eqref{e:continuity} in turn.  

\begin{enumerate}
\item 
We begin with Eq.~\eqref{e:continuity0}.
We have
\be
\lim_{x\rightarrow0^-}\hat Q(x,t) 
= -e^{2i\sigma_3\gamma}Q(0,t)
\ee
Thus, in order for the one-sided limits to coincide, we need either $Q(0,t)=0$ or $\gamma=\pi/2$.

\item 
Consider now Eq.~\eqref{e:continuity1}.
If the one-sided limits of $\hat Q(x,t)$ coincide at $x=0$ for all $t$, it follows that those for $\hat Q_t(x,t)$ will also coincide.  
So no further requirements arise.

\item 
Next, consider Eq.~\eqref{e:continuity2}.
We have
\be
\lim_{x\rightarrow0^-}\hat Q_x(x,t) 
= e^{2i\sigma_3\gamma}\left[Q_x(0,t)-ic\sigma_3Q(0,t)\right]
\ee
Thus, in order for the one-sided limits to coincide,
and taking into account the conditions we obtained when enforcing the continuity of $\hat Q(x,t)$ at $x=0$,
we need one of the following possibilities to hold:

\begin{enumerate}
\advance\itemsep2pt
\item 
$Q(0,t)=0$ and $\gamma=0$,
\item 
$Q_x(0,t)=(ic/2)\sigma_3 Q(0,t)$ and $\gamma=\pi/2$,
\end{enumerate}
The first and second possibilities yield respectively 
the extension~\eqref{e:Dext} and the homogeneous Dirichlet BCs~\eqref{e:DBC} or 
the extension~\eqref{e:Rext} and the homogeneous Robin BCs~\eqref{e:RBC}.
Note that a further possibility for the limits to coincide is that 
$Q(0,t)=0$ and $Q_x(0,t)=0$.
This possibility however is not consistent with the requirements for well-posedness of the BVP 
(which dictate that only one BC be imposed at the origin),
and therefore we discard it.  

\item 
Finally, consider Eq.~\eqref{e:continuity3}.
We have
\bml 
\label{e:continuity3a}
\lim_{x\rightarrow0^-}\hat Q_{xx}(x,t) 
= -e^{2i\sigma_3\gamma}[Q_{xx}(0,t)
\\
-c^2Q(0,t) -2ic\sigma_3Q_x(0,t)]\,.
\end{multline}
Taking into account the constraints already derived above,
we simply need to check that the equality is satisfied in the cases~(a) and~(b) above.
Indeed we have:
\begin{enumerate}
\advance\itemsep2pt
\item 
When $Q(0,t)=0$ and $\gamma=0$ (i.e., for homogeneous Dirichlet BCs)
\eqref{e:continuity3a} yields 
\be
Q_{xx}(0,t)=ic\sigma_3Q_x(0,t),
\label{e:continuity3b}
\ee
which would seem to impose a constraint on the solution.
Note however that when $Q(0,t)=0$, Eq.~\eqref{e:continuity3b} holds identically 
because $Q(x,t)$ satisfies the NLSLab equation.
\item 
When $Q_x(0,t)=(ic/2)\sigma_3Q(0,t)$ and $\gamma=\pi/2$  
[i.e., in the case of the special homogeneous Robin condition~\eqref{e:Rext}],
\eqref{e:continuity3a} is satisfied identically.
\end{enumerate}
\end{enumerate}


\subsection{Solution of the BVP}
\label{s:bvpsoln}

Having defined an IVP problem on the infinite line via the extension~\eqref{e:ext}, 
we can now proceed to solve the resulting BVPs using the formalism of section~\ref{s:IVP}.
All of the results of this section also reduce to those for the NLS equation~\eqref{e:nlslc} in the light-cone frame when $c=0$.  

One can define the Jost solutions as in section~\ref{s:IVP}
upon replacing $Q (x,t)$ by $\hat Q (x,t)$.  
With some effort one can show that this extension will lead to the following additional symmetry of the Jost functions 
by manipulating Eq.~\eqref{e:modjost}:
\bml
\mu^+ (x,t,k) \\= e^{i\sigma_3(\gamma+xc/2)}\mu^-(-x,t,-(k+c/2))e^{-i\sigma_3(\gamma+xc/2)}
\label{e:jostsym}
\end{multline}
In turn, inserting Eq.~\eqref{e:jostsym} into Eq.~\eqref{e:scat} 
one obtains the following symmetries for the scattering matrix: 
\be 
S^{-1}(-(k+c/2)) = e^{i\gamma\sigma_3}S(k)e^{-i\gamma\sigma_3}\,,\quad k\in\Real.
\ee
Combining this with the usual symmetry obtained from $Q^\dagger(x,t) = -Q(x,t)$, we also have 
\be
S^{-1}(-(k+c/2)) = S^\dagger(-(k+c/2)),\quad k\in\Real,
\ee
which implies the following relations for the scattering coefficients:
\bse
\begin{gather}
s_{11}^*(k) = s_{11}(-(k^*+c/2)),\quad \Im k\ge0,
\label{e:s11symm}
\\
s_{21}(-(k+c/2))=-e^{2i\gamma}s_{21}(k),\quad k\in\Real,
\end{gather}
\ese
where Eq.~\eqref{e:s11symm} was extended to the upper half plane using the Schwartz reflection principle.

The symmetry~\eqref{e:s11symm} implies that, 
if $k_j$ is an eigenvalue, so is $k_{j'}=-k_j^*-c/2$.
Thus, discrete eigenvalues come in symmetric quartets:
\be 
\left\{k_n,-k_n-c/2,k_n^*,-k_n^*-c/2\right\}_{n=1}^N,
\ee
with $N=J/2$, so there are $4N$ discrete eigenvalues symmetric about the point $-c/4$.    

The symmetries~\eqref{e:jostsym} of the eigenfunctions also induce corresponding symmetries for the norming constants.
In particular, denoting by $k_{n'} = - k_n^* - c/2$ the symmetric eigenvalue to $k_n$ and
by $c_n$ and $c_{n'}$ the associated norming constants,
we find 
$c_jc_{j'}^*=-e^{2i\gamma}$, 
which implies 
\be
C_jC_{j'}^* = e^{2i\gamma}/\left(\dot s_{11}(k_j)\right)^2
\label{e:normingsymmetry}
\ee
when $\gamma=n\pi/2$, 
where the dot denotes differentiation with respect to $k$
and $s_{11}(k)$ can be obtained from the trace formula.  
In particular, in the case $N=1$, and for reflectionless solutions,
\be 
C_1C_2^* 
= e^{2i\gamma}\left(\frac{(2k_1+c/2)(k_1-k_1^*)}{k_1+k_1^*+c/2}\right)^2.
\ee
Using the parametrization~\eqref{e:normingconst}, we also have 
\be 
\delta_j+\delta_{j'}=0,\hskip 20pt \psi_j-\psi_{j'}=2\gamma-\pi.
\ee

It should be noted that, for the NLS equation in the light cone frame, the full class of Robin BCs was linearized in \cite{Fokas1989} using 
a more complicated, but still explicit, ``energy''-dependent extension of the potential.
Such an extension, which was then also used in \cite{BH2009} to study the soliton behavior, introduces spurious singularities in the IST,
however, which must be carefully dealt with.
A ``cleaner'' and more elegant way to treat more general linearizable BCs is instead to use a suitable B\"acklund transformation.
We do so in section~\ref{s:BT}.

\section{Extending the potential via B\"acklund transformation}
\label{s:BT}

We now show how one can also extend the potential from the half line to the infinite line using a suitable B\"acklund transformation.
Similarly to what was done for the NLS equation in the light cone frame in 
\cite{JPA24p2507,IP7p435,SAM129p249},
this allows one to treat more general linearizable BCs than with the explicit extension discussed in section~\ref{s:extension}.

\subsection{The B\"acklund transformation}

Suppose that $q(x,t)$ and $\tilde q(x,t)$ are both solutions of the NLSLab equation and that their corresponding eigenfunctions $\phi(x,t,k)$ and $\tilde \phi(x,t,k)$ are related by the following transformation:
\be 
\tilde \phi(x,t,k)=B(x,t,k)\phi(x,t,k)\,.
\label{e:bt}
\ee
Following~\cite{JPA24p2507,IP7p435,SAM129p249},
we next show that, as a result, $q$ and $\tilde q$ are related by a B\"acklund transformation.  
To do so, we proceed as follows.
If $\phi(x,t,k)$ is invertible, we can use Eq.~\eqref{e:laxs} to obtain the following necessary and sufficient condition for Eq.~\eqref{e:bt} to hold:
\be 
B_x=-ik[\sigma_3,B]+\tilde QB-BQ.
\label{e:btode}
\ee
If $B(x,t,k)$ is linear in $k$, one can write a solution to Eq.~\eqref{e:btode} as
\be 
B(x,t,k) = -2ikI+\Lambda\sigma_3+d\sigma_3+(\tilde Q-Q)\sigma_3
\label{e:btsol}
\ee
where $\Lambda=\mathop{\rm diag}(\lambda_1,\lambda_2)$ is an arbitrary constant diagonal matrix and 
\be  
d(x,t) = \int_0^x\tilde q(y,t)\tilde r(y,t)-q(y,t)r(y,t)dy.
\label{e:dint}
\ee
A careful examination of the off-diagonal elements of Eq.~\eqref{e:btode} also yields the following transformation between $q$ and $\tilde q$:
\be 
\tilde q_x-q_x= (\lambda_2+d)\tilde q+(\lambda_1+d)q\,.
\label{e:qandqtilde}
\ee
Equation~\eqref{e:qandqtilde} is the desired B\"acklund transformation between the two solutions of the NLSLab equation.

For the NLS equation in the light cone frame, one then imposes the additional mirror symmetry $\tilde q(x,t) = q(-x,t)$, 
corresponding to the invariance $x\mapsto -x$ of the PDE.
As mentioned previously, the NLSLab equation does not possess the same symmetry as the standard NLS equation.  
However, we can impose a modified mirror symmetry between $q$ and $\tilde q$ as
\be 
\tilde q(x,t)=e^{icx}q(-x,t).
\label{e:mirrorsym}
\ee
Equation~\eqref{e:mirrorsym}, together with Eq.~\eqref{e:qandqtilde}, evaluated at $x=0$, yields the desired Robin BC
\be
q_x(0,t) + \alpha q(0,t) = 0
\label{e:RobinBCalpha}
\ee
when 
\be 
\frac{\lambda_1+\lambda_2}{2}=\alpha+\frac{ic}{2}.
\label{e:alphalambda}
\ee
Next we show that, like in the case of the standard NLS equation, however, there are constraints on values of $\alpha$ for which this construction is self-consistent.
To do so, use the B\"acklund transformation to solve the BVP, similarly to section~\ref{s:bvpsoln}.

\subsection{Solution of the BVP via B\"acklund transformation}

The approach to solve the BVP via the B\"acklund transformation is identical to that in Refs.~\cite{JPA24p2507,IP7p435,SAM129p249}.
Namely, given $q(x,0)$ for all $x>0$,
the transformation~\eqref{e:qandqtilde} and the condition~$\tilde q(0,0)=q(0,0)$ define $\tilde q(x,0)$ for all $x\ge 0$.  
Then we can use Eq.~\eqref{e:mirrorsym} to extend the original potential $q(x,0)$ to $x\le0$.  
This yields the extension
\be 
\hat q(x,t)=q(x,t)\Theta(x)+e^{icx}\tilde q(-x,t)\Theta(-x).
\label{e:BTextension}
\ee
One can then use the standard IST to solve the IVP for the extended potential~\eqref{e:BTextension} on the infinite line.
The restriction of the resulting solution to $x>0$ also solves the NLSLab on the half line with BCs~\eqref{e:linearizable},
similarly to section~\ref{s:bvpsoln}.
Next we study the properties of the IVP for the extended potential.

It is easy to show that, in the limit $\alpha\rightarrow \infty$, one has $\tilde q(x,t)=-q(x,t)$, which yields exactly the extension~\eqref{e:Dext}. 
Also, when $\alpha=-ic/2$, one obtains $\tilde q(x,t)=q(x,t)$, which yields exactly the extension~\eqref{e:Rext}.
Therefore we only need to discuss generic values of~$\alpha$.

Note that~\eqref{e:bt} relates \textit{generic} eigenfunctions of the corresponding Lax pairs.
Recalling the asymptotic behavior of the Jost solutions as $x\to\pm\infty$, 
we can obtain a relationship between the Jost solutions of the original Lax pair and the solutions of the B\"acklund-transformed counterpart as
\be 
\tilde\phi^{\pm}(x,t,k)=B(x,t,k)\phi^{\pm}(x,t,k)B_\infty^{-1}(k),
\label{e:btjost}
\ee
with 
\be
B_\infty(k)=\lim_{x\rightarrow\pm\infty} B(x,t,k) = -2ikI+\Lambda \sigma_3+d_\infty \sigma_3,
\ee
and $d_\infty = \lim_{x\rightarrow\pm\infty} d$.  
A simple calculation letting $y\mapsto -y$ in Eq.~\eqref{e:dint} and using Eq.~\eqref{e:mirrorsym} establishes that $\lim_{x\rightarrow\infty}d=\lim_{x\rightarrow-\infty}d$, so we also have $\lim_{x\rightarrow\infty}B=\lim_{x\rightarrow-\infty}B$.
Using Eq.~\eqref{e:btjost} together with the scattering relation~\eqref{e:scat}, we then obtain a corresponding relation for the scattering matrices as
\be 
\tilde S(k)=B_\infty(k)S(k)B_\infty^{-1}(k).
\ee

Similarly to section~\ref{s:bvpsoln}, we next derive the additional symmetries of the eigenfunctions
and scattering coefficients induced by the B\"acklund transformation.
Note first that the mirror symmetry~\eqref{e:mirrorsym} implies the existence of 
an additional relation between the Jost eigenfunctions of the original and transformed problem:
\be
\tilde \phi^{\pm}(x,t,k) = e^{icx\sigma_3/2}\sigma_3\phi^{\pm}(-x,t,-(k+c/2))\sigma_3.
\label{e:phitildepmrelation}
\ee
From Eq.~\eqref{e:phitildepmrelation} and the scattering relation~\eqref{e:scat}, we get, after some algebra, 
a corresponding relation between the scattering matrices of the original and transformed problem:
\be 
\tilde S(k)=\sigma_3S^{-1}(-(k+c/2))\sigma_3.
\ee
In turn, these relations, together with the relation~\eqref{e:btjost} between the Jost solutions, yield
the desired symmetry of the eigenfunctions and scattering matrix of the original problem:
\bml 
\phi^{\pm}(x,t,k)=B^{-1}(x,t,k)e^{icx\sigma_3/2}\sigma_3\\\times\phi^{\mp}(-x,t,-(k+c/2))\sigma_3B_\infty(k),
\label{e:jostefbt}
\end{multline}
\vglue-5ex
\be 
S(-(k+c/2))=\sigma_3B_\infty(k)S^{-1}(k)B_\infty^{-1}(k)\sigma_3,
\label{e:scattbt}
\ee
Combining Eq.~\eqref{e:scattbt} with the original symmetries of the scattering matrix~$S$, we obtain:
\bse
\begin{gather}
s_{11}^*(k) =s_{11}(-(k^*+c/2)),~~~~~
\label{e:s11bt}
\\
s_{21}(-(k+c/2)) =f(k)s_{21}(k),
\label{e:s21bt}
\\
\noalign{\noindent where}
f(k) =\frac{2ik+\lambda_2 + d_\infty}{2ik - \lambda_1- d_\infty}.
\end{gather}
\ese
In particular, Eq.~\eqref{e:s11bt} implies that discrete eigenvalues appear in symmetric quartets:
each discrete eigenvalue $k_j$ is associated to a symmetric eigenvalue
$k_{j'}=-(k_j^*+c/2)$,
in agreement with the results obtained through the explicit extension derived in section~\ref{s:extension}.

Up to this point, the constants $\lambda_1$, $\lambda_2$, and $d_\infty$ have remained arbitrary.  
On the other hand, it is easy to see that, applying Eq.~\eqref{e:s21bt} twice, one must have:
\be 
f(-(k+c/2))=1/f(k).
\ee
This restriction, along with Eq.~\eqref{e:alphalambda}, determines $\lambda_1$ and $\lambda_2$.
Explicitly,
\be
\lambda_1=\alpha\,,\qquad 
\lambda_2=\alpha+ic\,.
\label{e:lambda1lambda2}
\ee
We have thus completely determined the parameters of the B\"acklund transformation.

Next we use Eq.~\eqref{e:lambda1lambda2} to obtain the constraint on the value of~$\alpha$.
Recall that the scattering matrix is unitary, which implies
$|s_{11}(k)|^2+|s_{21}(k)|^2=1$.   
Obviously 
the same relation must hold when $k$ is replaced with $-(k+c/2)$.
Given the additional symmetry of the scattering coefficients arising from the B\"acklund transformation, however, 
this condition holds if and only if
\be 
|s_{11}(k)|^2+|f(k)|^2|s_{21}(k)|^2=1.
\ee
Thus, in order for the construction to be self-consistent, 
it must satisfy the requirement that $|f(k)|^2=1$.  
And it is straightforward to see that 
the constraint is only satisfied for
\be
\Im \alpha = -c/2\,.
\label{e:BCconstraint}
\ee
When $c=0$, this reduces exactly to the class of Robin BCs for the NLS equation in the light cone frame studied in Refs.~\cite{JPA24p2507,IP7p435,SAM129p249}.
Note also that, as in the case of the NLS equation in the light cone frame, the class of BCs identified by Eq.~\eqref{e:BCconstraint} reduces to 
the linearizable BCs that can be studied by an explicit extension when $\alpha = -ic/2$ and $\alpha=\infty$.


\begin{figure*}[t!]
\kern-\medskipamount
\centering \includegraphics[trim=0cm 0cm 0.5cm 0cm, clip=true,width=0.385\textwidth]{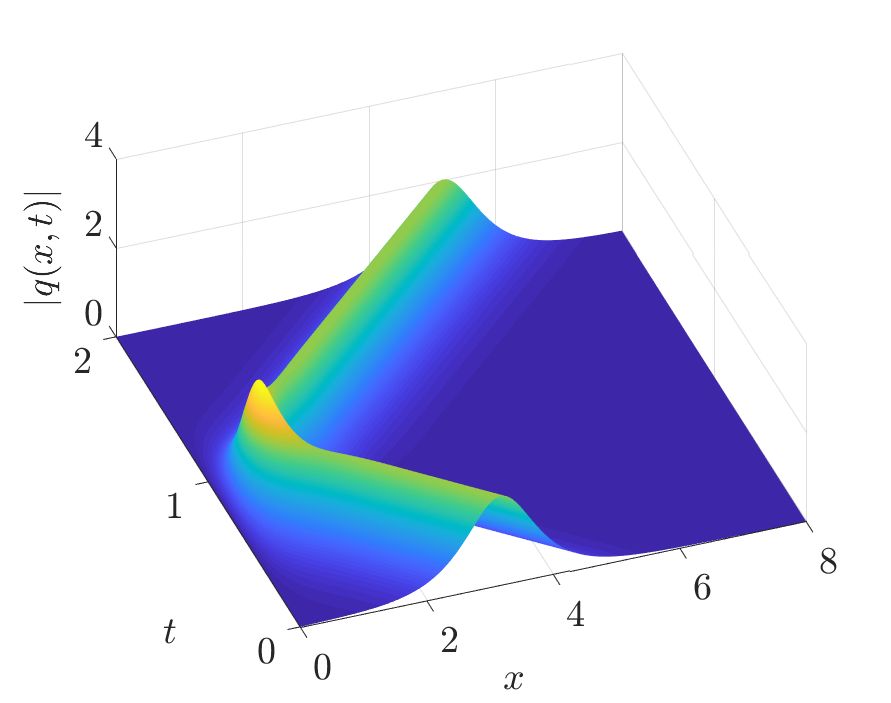}\qquad
\centering \includegraphics[trim=0cm 0cm 0.5cm 0cm, clip=true,width=0.360\textwidth]{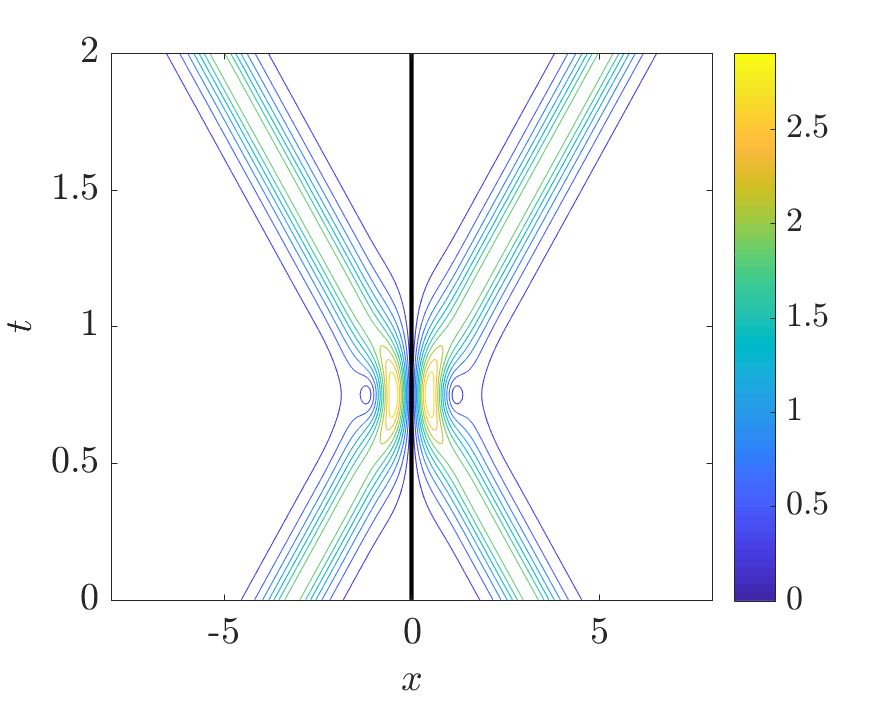}
\caption{A one-soliton solution reflected over the boundary $x=0$ in the laboratory frame in the case of Dirichlet BCs~\eqref{e:DBC}.  
Here $A = 2$, $V=-2.5$, $\delta = 3$ and $\psi = 0$.  
The left panel shows a three-dimensional plot of $|q(x,t)|$.
The right panel is a contour plot showing the physical soliton and its mirror.}
\label{fig:1solDBC}
\end{figure*}

\subsection{Norming constants and self-symmetric eigenvalues}

Although the above derivation of the B\"acklund transformation is complete, 
it will be useful to determine the value of $d_\infty$
when studying the behavior of the solutions of the BVP. 
We do this by using Eqs.~\eqref{e:jostefbt}, \eqref{e:scattbt}, and~\eqref{e:scat}, evaluated at $x=0$ and $k=-c/4$.  
From Eq.~\eqref{e:scattbt} at $k=-c/4$, we know that $s_{11}=s_{22}$ and $s_{12}=s_{21}=0$ when $f(-c/4)\ne1$.  
So we can conclude that, in this case, $S(-c/4)=s_{11}(-c/4)I$.  
We can also observe that $B(0,t,-c/4)=(\alpha+ic/2)\sigma_3$, 
so 
$B^{-1}(0,t,-c/4)=\sigma_3/(\alpha+ic/2)$.  
Finally, he have that $B_\infty(-c/4)=(\alpha + ic/2+d_\infty)\sigma_3$.  
Combining these relations, Eq.~\eqref{e:jostefbt} yields 
\be 
\left[s_{11}(-c/4)-\frac{\alpha + ic/2+d_\infty}{\alpha+ic/2}\right]\phi^-(0,t,-c/4)=0.
\ee
We then obtain
\be d_\infty = (s_{11}(-c/4)-1)(\alpha+ic/2).
\ee

It remains to determine the value of $s_{11}(-c/4)$.  
This can be done using the trace formula~\eqref{e:trace}.
When the values of $\alpha$ are restricted to ensure that $|f(k)|^2=1$, the integral in the right-hand side of Eq.~\eqref{e:trace} vanishes when $k=-c/4$.  
Let us refer to the discrete eigenvalues lying on the line $\Re(k_j)=-c/4$ as self-symmetric eigenvalues,
similarly to \cite{JPA24p2507},
and let us denote by 
$S$ the number of self-symmetric eigenvalues. 
Recalling~\eqref{e:BCconstraint} and the trace formula~\eqref{e:trace} at $ k = -c/4$ yields
\be 
d_\infty = ((-1)^S-1)\,\Re\alpha\,.
\ee

The final piece of information needed is the symmetry between the norming constants associated to symmetric eigenvalues.  
This, however, can easily be done using Eq.~\eqref{e:jostefbt}, obtaining 
\bse
\be 
c_jc_{j'}^*=\frac{1}{f(k_j)}.
\ee
Or, equivalently,
\be
C_jC_{j'}^*=-\frac{1}{f(k_j)(\dot s_{11}(k_j))^2}. 
\ee
\ese
In the special cases of $\alpha = -ic/2$ and the limit $\alpha=\infty$, this equation reduces to Eq.~\eqref{e:normingsymmetry} 
for $\gamma = \pi/2$ and $\gamma = 0$, respectively.
The resulting behavior for the solutions is discussed in section~\ref{s:solitons}.

Note that, from a rigorous point of view, 
when using the B\"acklund transformation to extend the potential,
one should make sure that the extension is still in the class of potentials that can be treated by the IST on the infinite line.
In practice, this requires proving that if $q(x,0)\in L^1(\Real^+$) the extended potential $\hat q(x,0)$ is in $L^1(\Real)$.
Such a proof is nontrivial, however, and is therefore outside the scope of this work.
For the NLS equation in the light cone frame, such a proof can be found for example in \cite{deiftpark}.

\begin{figure*}[t!]
\kern-\medskipamount
\centering \includegraphics[trim=0cm 0cm 0.5cm 0cm, clip=true,width=0.385\textwidth]{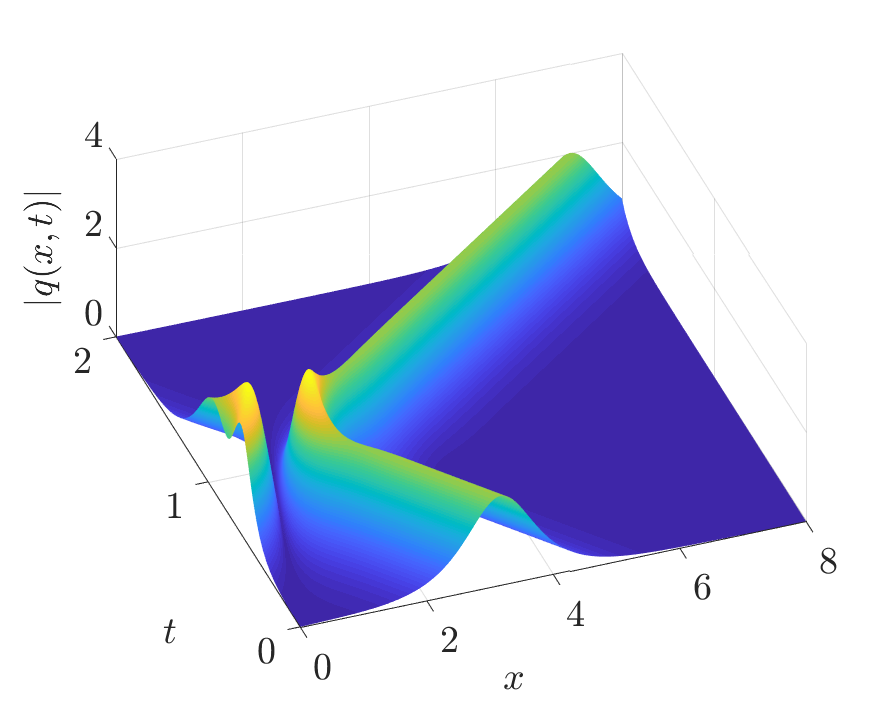}\qquad
\centering \includegraphics[trim=0cm 0cm 0.5cm 0cm, clip=true,width=0.360\textwidth]{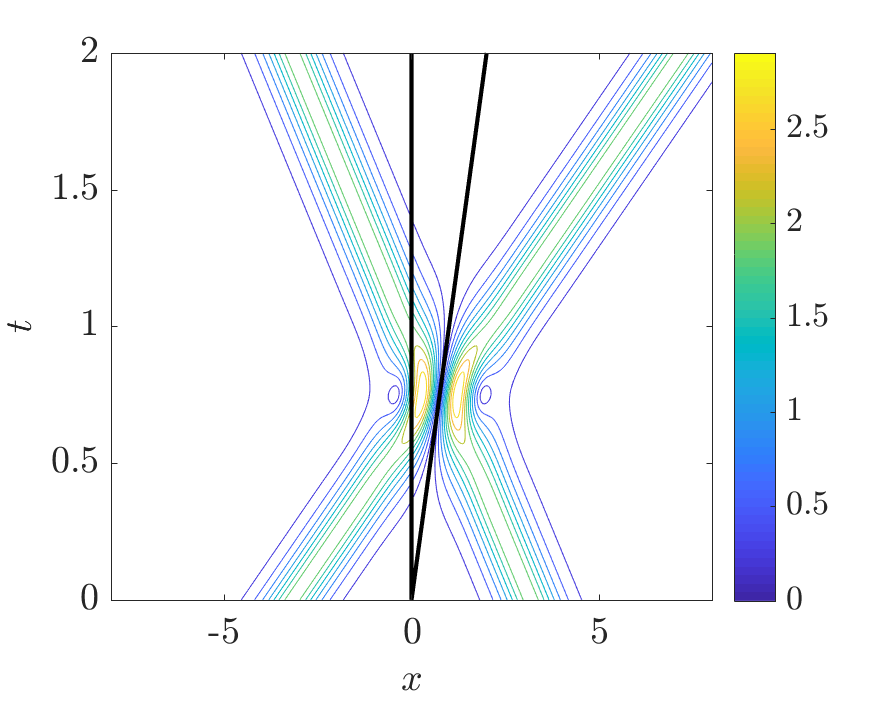}
\caption{Same as in Fig.~\ref{fig:1solDBC},
but for a one-soliton solution with Dirichlet BCs imposed at $X=0$ in the light cone frame 
(i.e., over the physical boundary $x=ct$ in the laboratory frame), 
here shown in the laboratory frame.  
Here $A = 2$, $V=-2$, $\delta = 3$ and $\psi = 0$.  
The black line in the right panel is the line $x=ct$ (i.e., $X=0$).}
\label{fig:1solLCinLab}
\medskip
\centering \includegraphics[trim=0cm 0cm 0.5cm 0cm, clip=true,width=0.385\textwidth]{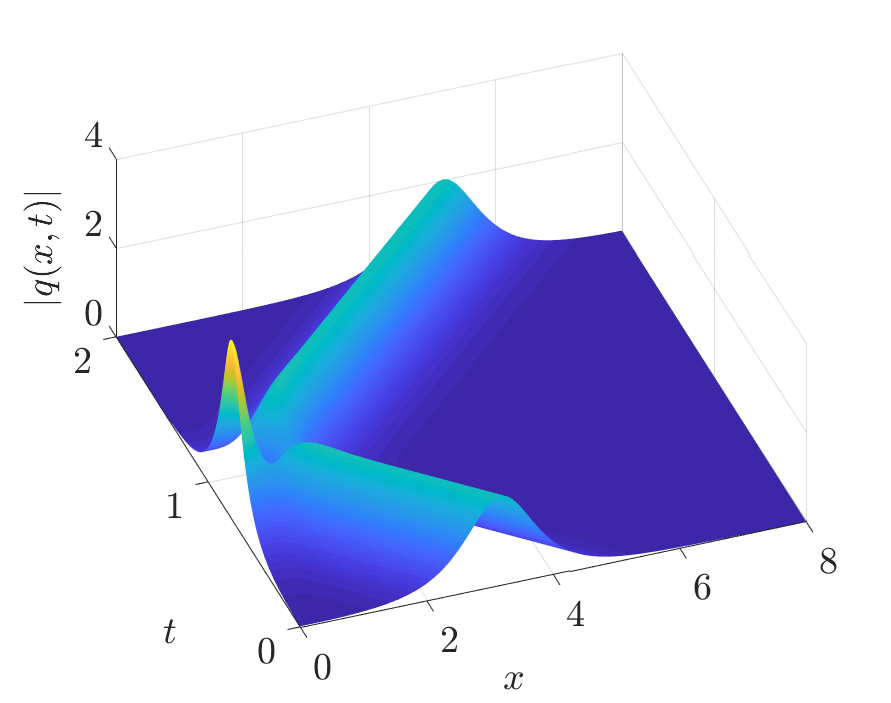}\qquad
\centering \includegraphics[trim=0cm 0cm 0.5cm 0cm, clip=true,width=0.360\textwidth]{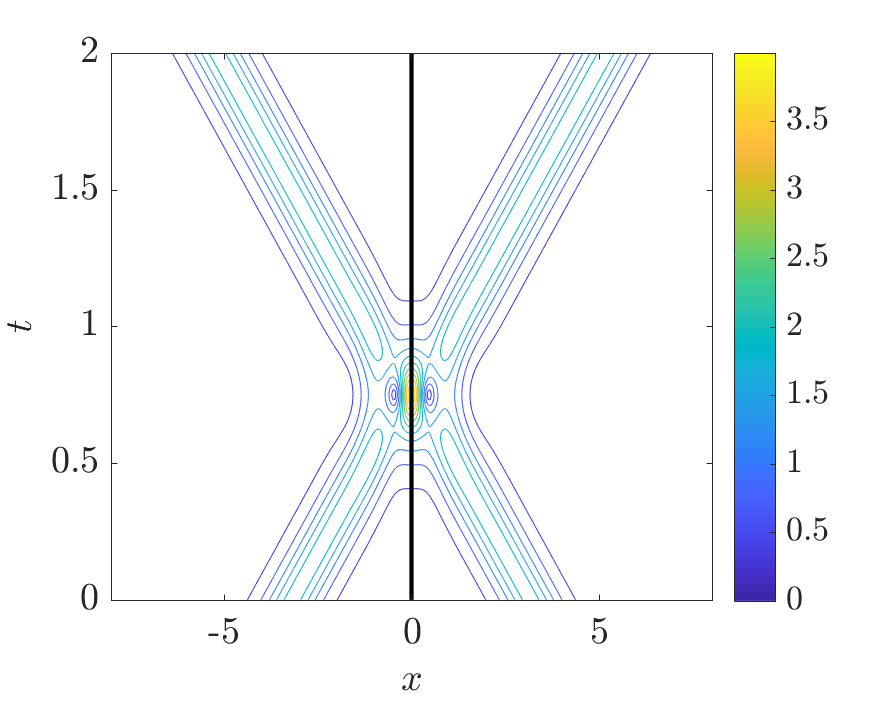}
\caption{Same as Fig.~\ref{fig:1solDBC}, but for the special Robin BCs~\eqref{e:RBC}.  
Here $A = 2$, $V=-2.5$, $\delta = 3$ and $\psi = 0$.}
\label{fig:1solRBC}
\end{figure*}

\begin{figure*}[t!]
\kern-\medskipamount
\centering \includegraphics[trim=0cm 0cm 0.5cm 0cm, clip=true,width=0.385\textwidth]{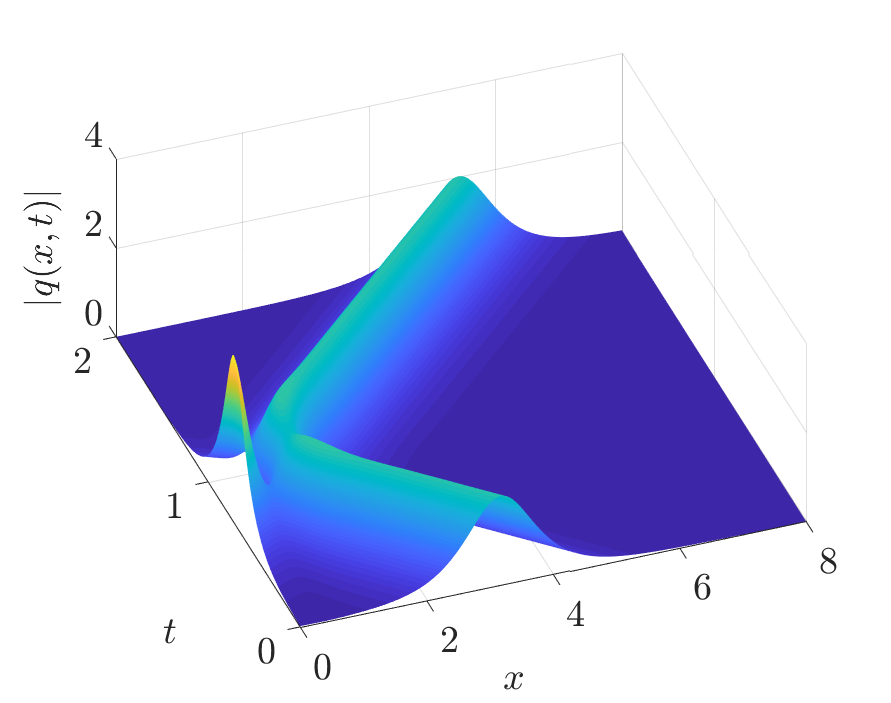}\qquad
\centering \includegraphics[trim=0cm 0cm 0.5cm 0cm, clip=true,width=0.360\textwidth]{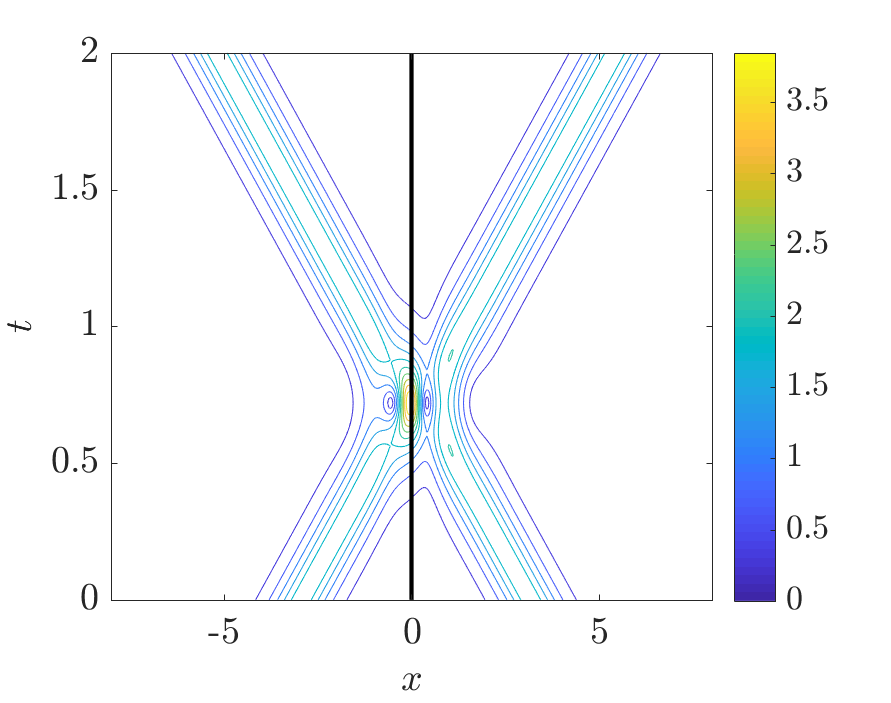}
\caption{Same as Fig.~\ref{fig:1solRBC}, but for a different kind of Robin BCs, namely Eq.~\eqref{e:RobinBCalpha} with $\alpha = 1-ic/2$.  
Here $A = 2$, $V=-2.5$, $\delta = 3$ and $\psi = 0$.
}
\label{fig:1solRBC2}
\medskip
\centering \includegraphics[trim=0cm 0cm 0.5cm 0cm, clip=true,width=0.385\textwidth]{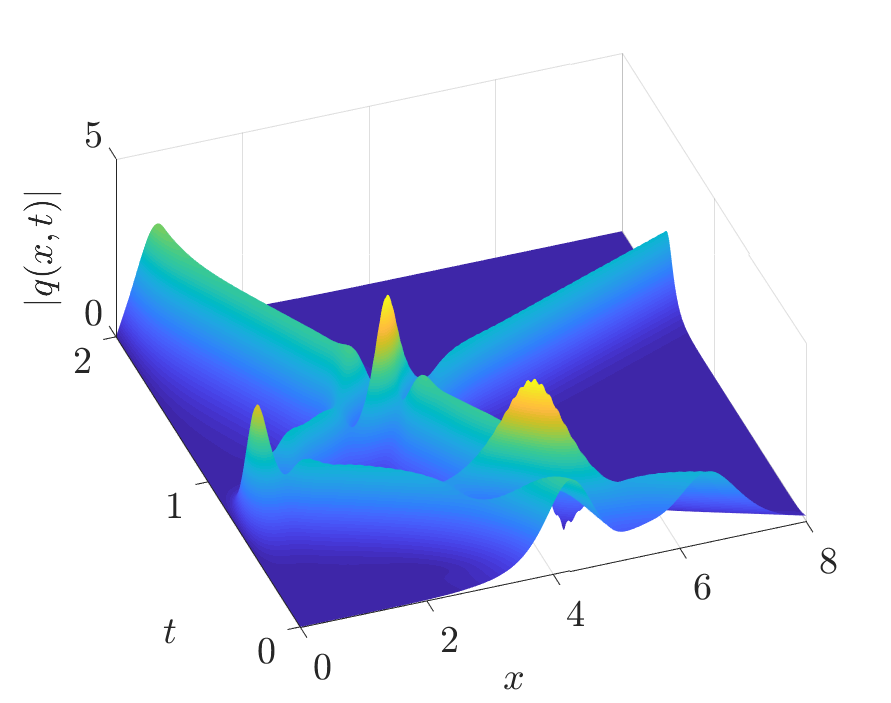}\qquad
\centering \includegraphics[trim=0cm 0cm 0.5cm 0cm, clip=true,width=0.360\textwidth]{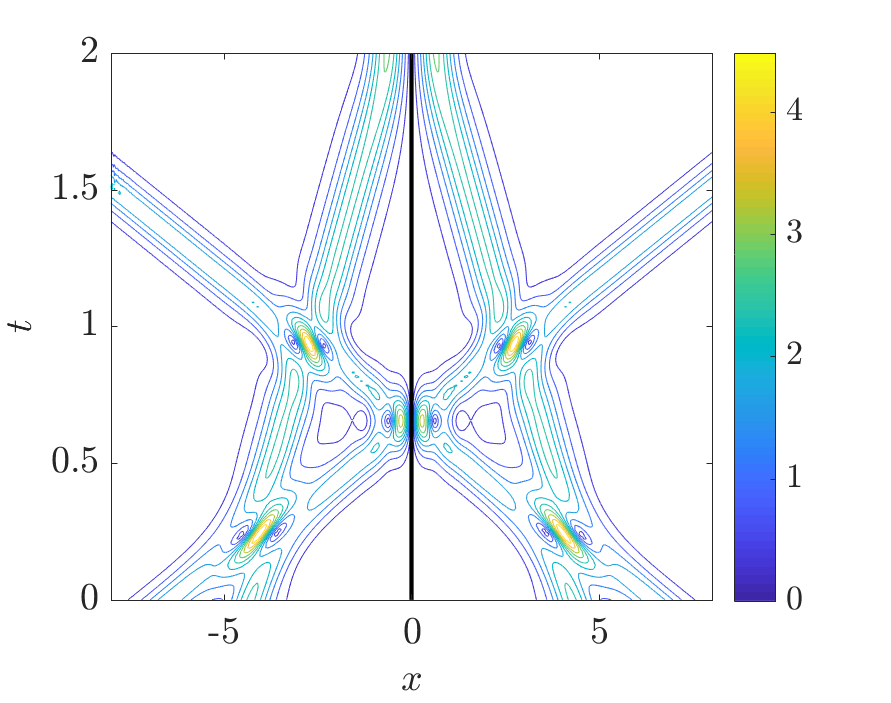}
\caption{Two solitons reflected over the boundary with Dirichlet BCs~\eqref{e:DBC}.  
Here $A_1 = 2$, $A_2=2.5$, $V_1=-5$, $V_2=-1.5$, $\delta_1 = 6$, $\delta_2=4$ and $\psi_1 = \psi_2=0$.
}
\label{fig:2solDBC}
\medskip
\centering \includegraphics[trim=0cm 0cm 0.5cm 0cm, clip=true,width=0.385\textwidth]{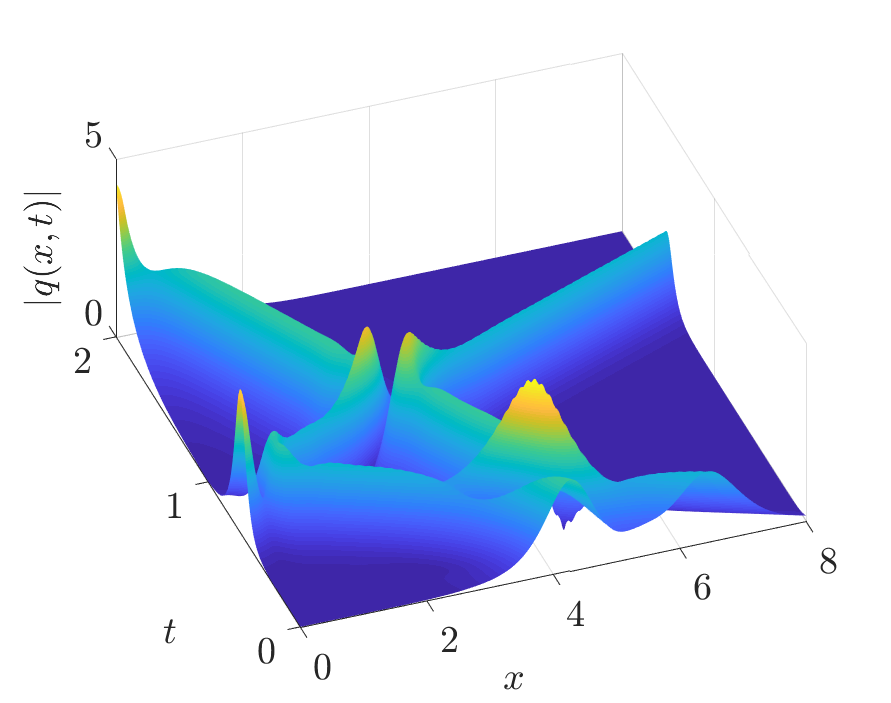}\qquad
\centering \includegraphics[trim=0cm 0cm 0.5cm 0cm, clip=true,width=0.360\textwidth]{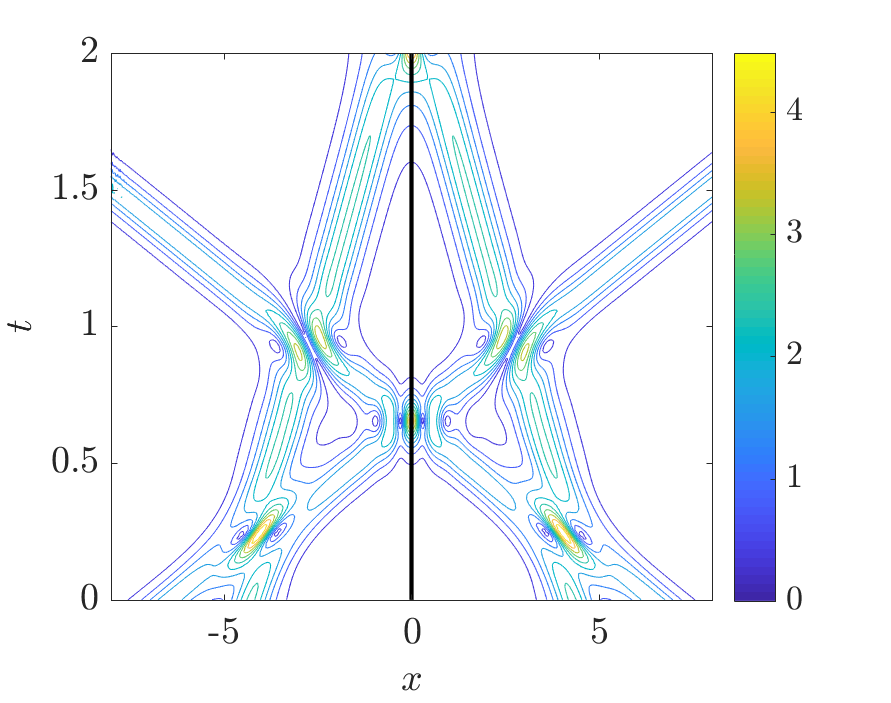}
\caption{Two solitons reflected over the boundary in the case of the special Robin BCs~\eqref{e:RBC}.  
The soliton parameters are the same as in Fig.~\ref{fig:2solDBC}.
}
\label{fig:2solRBC}
\end{figure*}

\section{Solution behavior}
\label{s:solitons}

Although the solution of the BVP in sections~\ref{s:extension} and~\ref{s:BT} is general,
for simplicity in what follows we will only look at reflectionless potentials, i.e., pure soliton solutions.

Figure~\ref{fig:1solDBC} shows a one-soliton solution of the BVP for the NLSLab equation in the case of Dirichlet BCs.
As in previous studies \cite{BH2009}, the soliton appears to reflect and bounce back off the boundary in the left panel.  
In terms of the extension to the infinite line presented in section~\ref{s:bvpsoln},
what is actually happening is perhaps more clear in the contour plot on the right: 
the physical soliton passes through the boundary,
while at the same time a mirror soliton that came through the boundary with equal amplitude and opposite velocity 
takes over and continues propagating in the physical domain.  
At first sight, this behavior appears the same as in the solutions of the light cone NLS equation presented in \cite{BH2009}, 
except for the change in the reference frame.  
A crucial difference exists, however.
In the solution of the BVP for the NLS equation in the light cone,
the mirror solitons had an with equal and opposite velocity in that reference frame.  
Recall that the one-soliton solution of the NLS equation~\eqref{e:nlslc} generated by a discrete eigenvalues $k=(V+iA)/2$
is given by Eq.~\eqref{e:nlslc1soliton},
which travel in that frame at group velocity $2V$.  
But since the reference frame moves at velocity $c$ with respect to the laboratory frame, 
the actual velocity of the soliton in the laboratory frame is $2V+c$.  
Thus, a reflected mirror soliton that travels with velocity $-2V$ in the light cone frame 
actually has physical velocity $-2V+c$.  
Hence, two solitons reflected over the physical boundary $x=ct$ do not have opposite group velocities
in the laboratory frame.   
To illustrate this fact, Fig.~\ref{fig:1solLCinLab} shows a one-soliton solution with homogeneous Dirichlet BCs at $X=0$ 
(i.e., $x=ct$ in the laboratory frame),
shown in the laboratory frame.  
It is clear from the figure that the actual velocities of the solitons in the latter are not equal and opposite.

In contrast, in the NLSLab equation, a soliton generated by a discrete eigenvalue $k=(V+iA)/2$ is given by~\eqref{e:soliton},
which travels with group velocity $2V+c$, while the reflected mirror solitons travels with group velocity $-(2V+c)$.  
Thus, two solitons reflected over the physical boundary $x=0$ of the laboratory frame do have opposite physical group velocities, 
as shown in Fig.~\ref{fig:1solDBC}.  

Similar results apply to BVPs with Robin BCs.  
To illustrate this, Fig.~\ref{fig:1solRBC} shows a one-soliton solution and its reflection at the boundary 
in the case of the Robin BCs~\eqref{e:RBC},
and Fig.~\ref{fig:1solRBC2} shows a solution for a special case of the Robin BCs~\eqref{e:BCconstraint}.
Moreover, it should be clear that the methods
presented in sections~\ref{s:bvpsoln} and~\ref{s:BT} do not only apply to one-soliton solutions.
To demonstrate this, Figs.~\ref{fig:2solDBC} and~\ref{fig:2solRBC} 
show two two-soliton solutions and their reflections at the boundary, 
first in the case of Dirichlet BCs and in the case of the Robin BCs~\eqref{e:RBC}.

\section{Conclusions}
\label{s:conclusions}

We studied IVPs and BVPs for the NLS equation in the lab frame.
In particular, we constructed a modified extension from the half line to the infinite line
which allowed us to identify a class of linearizable BCs for the BVP on the half plane,
comprised of homogeneous Dirichlet BCs and a special type of homogeneous Robin BCs.
We have also shown how BVPs for the NLSLab equation on the half line with these BCs can be solved using this extension.

The IST for the NLSLab equation on the infinite line is essentially the same as in the light-cone regime, 
except for an additional term in the time dependence.  
In particular, a soliton corresponding to a discrete eigenvalue $k=(V+iA)/2$ has amplitude $A$ (as for the standard NLS equation) but travels with velocity $2V+c$.  
In the BVP, we have shown that, as for the standard NLS equation, a symmetry exists between the discrete eigenvalues of the scattering problem.
For the NLSLab equation, however, the symmetry is different:
a soliton corresponding to a discrete eigenvalue $k_1=(V+iA)/2$ has a mirror soliton 
corresponding to the symmetric eigenvalue $k_2=-(V+c+iA)/2$.
On the other hand, while the symmetry relation is different, the relation between the physical properties of a solitons and its mirror
are the same for the standard NLS equation and the NLSLab equation, since in both cases the mirror soliton 
has the same amplitude but opposite velocity to that of the ``physical'' soliton.  

From a physical point of view it should be noted that the derivation of the NLS equation as a model for the propagation of quasi-monochromatic
optical pulses is done under the framework of unidirectional propagation.  
Thus, the validity of the NLS equation as a model to describe pulses that propagate backwards with sufficient velocity to hit the 
boundary of the physical domain $x=0$ is certainly questionable.

On the other hand, the results of this Letter raise an interesting question from a mathematical point of view. 
It is generally thought that the linearizable BCs are those BCs that preserve all of the symmetries of an integrable evolution equation.
The fact that the class of linearizable BCs for the NLS equation in the laboratory frame differs 
from that for the NLS equation in the light-cone frame, however,
shows that, perhaps surprisingly, the class of lineariable BCs is different for each equation in the NLS hierarchy.
This is because the NLS equation in the laboratory frame is simply the flow obtained from the combination of 
the NLS equation in the light cone frame and the one-directional wave equation, which is just the preceding member 
of the NLS hierarchy.
Since all equations in the NLS hierarchy possess the commuting flows, this raises the question of 
how precisely the class of linearizable BCs is related to the second half of the Lax pair and to the symmetries of the evolution equation.
We hope the results of the present work will stimulate further work on this question.

\medskip
\textit{Acknowledgements.}
KL is grateful to M. Schwarz for many helpful discussions.
This work was partially supported by the National Science Foundation under grant number DMS-1615524.

\section*{References}
 
\catcode`\@ 11
\def\journal#1&#2,#3 (#4){\begingroup \let\journal=\d@mmyjournal {\frenchspacing\sl #1\/\unskip\,} {\bf\ignorespaces #2}\rm,\ #3 (#4)\endgroup}
\def\d@mmyjournal{\errmessage{Reference foul up: nested \journal macros}}
\def\@biblabel#1{#1.}

\end{document}